\documentclass{article}
\usepackage{spconf,amsmath,graphicx,booktabs,siunitx}

\title{Analysis of the BUT Diarization System for VoxConverse Challenge}
\name{\begin{tabular}{c}Federico Landini, Ond\v{r}ej Glembek, Pavel Mat\v{e}jka, Johan Rohdin, \\ Luk\'{a}\v{s} Burget, Mireia Diez, Anna Silnova \thanks{The work was supported by Czech National Science Foundation (GACR) project ``NEUREM3'' No. 19-26934X, European Union’s Horizon 2020 project No. 833635 ROXANNE, Czech Ministry of Interior projects No. VI20192022169 ``AI v TiV'' and No. VJ01010108 ``ROZKAZ'', and Czech Ministry of Education, Youth and Sports project No. LTAIN19087 ``Multi-linguality in speech technologies''.}\end{tabular}}
\address{Brno University of Technology, Faculty of Information Technology, Speech@FIT, Czechia}
\usepackage{xcolor}

\usepackage[framemethod=tikz]{mdframed}
\usepackage{tikz} 
\usepackage{graphicx}
\usepackage[colorinlistoftodos]{todonotes}
\usepackage[colorlinks=true, allcolors=blue]{hyperref}
\usepackage{enumitem}
\usepackage{verbatim} 

\begin{document}
\ninept
\maketitle
\begin{abstract}
This paper describes the system developed by the BUT team for the fourth track of the VoxCeleb Speaker Recognition Challenge, focusing on diarization on the VoxConverse dataset. The system consists of signal pre-processing, voice activity detection, speaker embedding extraction, an initial agglomerative hierarchical clustering followed by diarization using a Bayesian hidden Markov model, a reclustering step based on per-speaker global embeddings and overlapped speech detection and handling. We provide comparisons for each of the steps and share the implementation of the most relevant modules of our system. 
Our system scored second in the challenge in terms of the primary metric (diarization error rate) and first according to the secondary metric (Jaccard error rate).
\end{abstract}
\begin{keywords}
Speaker Diarization, Variational Bayes, HMM, VoxConverse, VoxSRC Challenge
\end{keywords}

\section{Introduction}
\vspace{-2mm}
\label{sec:intro}
Speaker diarization applied to broadcast data has been of interest for decades in part due to the potential applications such as speech collection of a speaker of interest, speech search, segmentation or automatic transcription. In order to address the lower performance of systems on data from radio, television or web videos, during the last decade, new datasets allowing diarization on such types of data were released: ETAPE~\cite{gravier2012etape}, MGB-1~\cite{bell2015mgb}, Albayzin~\cite{lleida2019albayzin} and to a lesser extent DIHARD~\cite{ryant2019second}. Although participants of the challenges related to the datasets, can access the data without fees, any other party interested in evaluating their system on such corpora needs to pay a fee.

VoxConverse~\cite{chung2020spot} is a dataset of videos `in the wild' collected from YouTube consisting in talk shows, news broadcasts, celebrity interviews, home vlogs, etc. which is released publicly and for free\footnote{\url{http://www.robots.ox.ac.uk/~vgg/data/voxconverse}}. The organizers of the VoxSRC Challenge, famous for presenting a benchmarking dataset for speaker recognition on VoxCeleb~\cite{chung2018voxceleb2} proposed a fourth track in VoxSrc Challenge 2020 focused on audio diarization on VoxConverse.

In this paper we present the system devised by Brno University of Technology for the challenge with the corresponding analyses of results on the development set of the corpus. 
Furthermore, we identify the main challenges with the dataset and where the focus should be put on to improve the performance of diarization systems.

Although the proposed system is built on existing technologies, we try for the first time in our pipeline a voice activity detection (VAD) module combining different systems. However, our main contribution is the analysis of performance of each of the modules that compose our best system.
Such system was tuned (and not trained) on the development set of VoxConverse and reached a diarization error rate (DER) of 4\% and a Jaccard error rate (JER) of 19.8\%. Such DER represents almost half of the error reported with the baseline system~\cite{chung2020spot} when both audio and visual cues were used for performing diarization. The submission of this system for the evaluation set obtained the second position in the challenge in terms of the primary evaluation metric (8.12\% DER) and the first position according to the secondary metric (18.35\% JER). Together with this publication we make available the most relevant modules of our system~\cite{VBx}.

\vspace{-1mm}
\section{SYSTEM OVERVIEW}
\vspace{-2mm}
Our diarization system comprises the following steps:
\begin{enumerate}
\itemsep-0.25em
    \item Signal pre-processing
    \item Voice activity detection
    \item Speaker embedding extraction
    \item Initial clustering
    \item Variational Bayes hidden Markov model (HMM) clustering
    \item Global speaker embedding reclustering
    \item Overlap speech detection and handling
\end{enumerate}

In the following subsections, we describe each of the steps and present comparative results to understand the improvement provided by these steps. The diarization performance will be presented in terms of DER, its three components: missed speech, false alarm speech, and speaker error, and JER. Throughout the paper all diarization results are obtained with 0.25s forgiveness collar for DER~\cite{rt09} following the challenge protocol while for JER~\cite{ryant2019second} there is no collar by definition. VAD and overlapped speech detection (OVD) systems are evaluated without forgiveness collar as well.

\vspace{-2mm}
\subsection{Signal pre-processing}
\vspace{-1mm}
We considered two methods for signal preprocessing: the speech enhancement method based on a long short-term memory (LSTM) network trained on simulated data~\cite{sun2018speaker} (also used in the baseline) and the weighted prediction error (WPE)~\cite{nakatani2010speech, drude2018nara} as it had proved to be useful in the Second DIHARD challenge~\cite{landini2020but}. In our experiments, we saw that using the LSTM-based speech enhancer was beneficial while the WPE method was actually harmful. Since the latter removes late reverberation, and many of the recordings in VoxConverse are captured with studio microphones in close-distance, it is not surprising that this method does not help with this type of recordings. 
Therefore, LSTM based enhancement was applied in all experiments shown below.

\vspace{-2mm}
\subsection{Voice activity detection}
\vspace{-1mm}
\label{sec:VAD}
One key aspect of the challenge is that there are no ground truth VAD labels to use. In our pipeline, this means that labels have to be produced automatically before extracting embeddings. For this purpose, we evaluated three systems: 
\begin{itemize}[leftmargin=5mm]
    \item an energy-based VAD.
    \item a deep neural network (DNN) based system with three feed-forward layers receiving as input $\pm$5 stacked frames and trained to output 10ms frame decisions (silence / speech)~\cite{landini2020but}. It was trained on part of the second DIHARD development set (the rest was used for validation while training), the train set of the ``full-corpus'' partition of AMI\footnote{\url{http://groups.inf.ed.ac.uk/ami/corpus/datasets.shtml}}~\cite{carletta2005ami} (the test and development sets were used for validation while training), ICSI~\cite{janin2003icsi} and ISL~\cite{burger2002isl} meetings.
    \item an automatic speech recognition (ASR) based system. The frame-level phoneme labels were generated using the official Kaldi \cite{povey2011kaldi} Tedlium speech recognition recipe ({\texttt{s5\_r3}}) based on the TED-LIUM 3 dataset~\cite{hernandez2018ted}. 
    Phoneme classes corresponding to silence and noise were considered silence for the purpose of VAD and the rest of the classes were considered speech.
\end{itemize}

For the energy and DNN based systems a median filter was applied to ``smooth'' the outputs. For both models the detection threshold and span of the median filter window were optimized (individually) so that the sum of false alarm (FA) and miss, comprising the total error, were minimized on VoxConverse development set. 

When analyzing the outputs given by the systems we found out that in many cases speech segments were separated by short periods of silence (especially on the ASR based one). We therefore labeled segments of silence shorter than a certain length as speech in order to decrease the total error. Note that this is not equivalent to using a more tolerant threshold as the post-processing only affects short pauses.

In order to leverage the performance of individual systems, we used the outputs of the three models before removing short segments of silence in a majority voting system. Then, we removed silences shorter than 0.6s to improve the performance further. Table~\ref{tab:VAD_performance} presents the performance of each method in terms of different metrics. We see that all the methods evaluated surpass the performance of the baseline VAD. Surprisingly, the energy-based system performs quite well in comparison with more sophisticated methods. Still, it should be noted that the DNN-based method was trained mostly on meeting-like recordings. Even if DIHARD contains a more diverse set of recordings, they are not necessarily mostly broadcast like VoxConverse. Most likely, by making use of annotated multi-media recordings like the ones used in the challenge, the performance could still be improved. Analogously, using an ASR system trained on matched data could attain better performance.

Finally, taking advantage of the high precision of the ASR based system, we improved the overall performance by marking as silence any segment appearing more than 0.8s afar from speech detected by the ASR system. We used this VAD for the following results.

\begin{table}[!tb]
  \centering
  \caption[dummy caption to avoid error with footnote]{Frame accuracy, precision, recall, miss, false alarm, and total error for different VADs on the development set. WebRTC was the method used in the baseline~\cite{chung2020spot} (VAD annotations taken from the diarization output\setcounter{footnote}{3}\footnotemark).}
  \label{tab:VAD_performance}
  \vspace{2mm}
  \setlength{\tabcolsep}{4pt} 
  \begin{tabular}{@{}
                  l
                  S[table-format=2.2]  
                  S[table-format=1.3]  
                  S[table-format=1.3] 
                  S[table-format=2.2]  
                  S[table-format=1.2]  
                  S[table-format=2.2] 
                  @{}}
  \toprule
        VAD       & 
        \multicolumn{1}{c}{Acc.} & 
        \multicolumn{1}{c}{Prec.} & 
        \multicolumn{1}{c}{Reca.} &
        \multicolumn{1}{c}{Miss} & 
        \multicolumn{1}{c}{FA} & 
        \multicolumn{1}{c}{Error} \\
  \midrule
  WebRTC & 87.95 & 0.952 & 0.914 & 7.81 & 4.24 & 12.05 \\
  \midrule
  Energy & 94.56 & 0.953 & 0.989 & 1.01 & 4.44 & 5.45 \\
  - sil $<$ 0.3s & 94.66 & 0.952 & 0.992 & 0.77 & 4.57 & 5.34 \\
  \midrule
  DNN & 96.24 & 0.978 & 0.981 & 1.71 & 2.05 & 3.76 \\
  - sil $<$ 0.7s & 96.37 & 0.974 & 0.986 & 1.24 & 2.39 & 3.63 \\ 
  \midrule
  ASR & 83.30 & 0.989 & 0.826 & 15.84 & 0.86 & 16.70\\ 
  - sil $<$ 1.1s & 96.28 & 0.973 & 0.987 & 1.21 & 2.51 & 3.72 \\ 
  \midrule
  Majority voting & 96.62 & 0.978 & 0.985 & 1.37 & 2.01 & 3.38 \\ 
  - sil $<$ 0.6s & 96.82 & 0.975 & 0.990 & 0.87 & 2.31 & 3.18 \\ 
  - dist. ASR $<$ 0.8s & 97.02 & 0.978 & 0.989 & 0.97 & 2.01 & 2.98 \\ 
  \bottomrule
  \end{tabular}
\end{table}
\footnotetext{\url{https://github.com/a-nagrani/VoxSRC2020/blob/master/data/diar/baseline.rttm}}

\vspace{-2mm}
\subsection{Speaker embeddings}
\vspace{-1mm}
As explained above, a key element in the diarization pipeline is the extraction of x-vectors. In our system, 256 dimensional speaker embeddings were extracted from a 152-layer ResNet~\cite{he2016resnet} DNN.  The network was trained on the development part of the VoxCeleb~2 dataset~\cite{chung2018voxceleb2} (5994 speakers in ~145k sessions), cut into 2-second chunks and augmented with noise, as described in~\cite{snyder2018Xvectors} and available as part of the Kaldi-recipe collection~\cite{povey2011kaldi}. As input, we used 64-dimensional filter-banks extracted from the original 16\,kHz audio with a window size of 25\,ms and a 10\,ms shift.  

The loss function used for training the DNN was CosFace~\cite{want2018cosface}, with scaling parameter $s$ set to 32 and margin parameter $m$ linearly increased from $0.05$ to $0.3$ throughout the whole period of training.  We ran 1 epoch (i.e., passing all training data once) of stochastic gradient descent optimization, throughout which we exponentially decreased the learning rate from $10^{-1}$ to $10^{-6}$.  Note that we scaled the learning rate by the number of parallel jobs to compensate for the dynamic range of the accumulated gradients, in our case by 3.  The momentum and weight decay were kept constant at $0.9$ and $5 \cdot 10^{-4}$, respectively.  The batch size was set to 128, however, training on 3 GPUs in parallel virtually tripled the batch size.  Also, due to large memory requirements, the gradients were computed over 2 ``micro-batches'' of size 64 after which the update step was taken.  Note that care needs to be taken when using this approach in connection with any model that uses batch-normalization (as our ResNet model does) as batch-norm statistics may get biased with decreasing batch size.

\vspace{-2mm}
\subsection{Initial clustering}
\vspace{-1mm}
As the Bayesian HMM requires an initial assignment of frames to speakers, one possibility is to assign them randomly to a set of speakers. However, the model benefits from using a more sensible initial assignment. As in previous work~\cite{landini2020but}, the  x-vectors extracted from an input recording are clustered by means of agglomerative hierarchical clustering (AHC) with similarity metric based on probabilistic linear discriminant analysis (PLDA)~\cite{kenny10PLDA_HTP} log-likelihood ratio scores, as used for speaker verification. The PLDA  model for this  purpose was trained on x-vectors extracted from concatenated speech segments from VoxCeleb 2~\cite{chung2018voxceleb2} which are mean-centered, whitened to have identity covariance matrix and length-normalized~\cite{GarciaRomero2011lnorm}. Unlike in the baseline, x-vectors are extracted on 1.5\,s seconds segments but with overlap of 1.25\,s (instead of 0.75\,s) as this proved to be beneficial~\cite{diez2020optimizing}. When applying principal component analysis (PCA) per recording to project the embeddings to few dimensions, we found that keeping 55\% of the variability provided better performance than when using only 30\% as in \cite{diez2020optimizing} or when using 10\% as in the baseline. This could be explained by the fact that we have better embeddings than before which encode more relevant information for the task. 

The AHC clustering threshold was tuned on the development set to perform well in tandem with the next step as slightly underclustering allows for better performance when combined with variational Bayes (VB) HMM diarization. A comparison of the performance when using the threshold to undercluster 
and the one that minimizes DER 
is presented in Table~\ref{tab:VB_performance} rows 2 and 4 respectively. It is clear that a great difference in performance with regard to the baseline comes from an improved VAD; however, improved speaker embeddings and more frequent segmentation explain the rest of the difference reaching more than 76\% relative improvement in terms of DER and more than 58\% in terms of JER.

\vspace{-2mm}
\subsection{Bayesian HMM for x-vector clustering}
\vspace{-1mm}
Some of the problems of employing AHC for performing diarization are that it heavily depends on correctly tuning a clustering threshold for having good performance and that, it does not make use of the temporal nature of the embeddings when assigning them to speakers. A more principled way of addressing diarization is to model the problem with a Bayesian HMM where the complexity control of the Bayesian learning allows to infer the number of speakers and where the HMM transitions naturally model time dependencies.
In our Bayesian HMM model, the HMM states represent speakers, the transition between states represent the speaker turns and the state distributions are derived from a PLDA model pre-trained on labeled x-vectors in order to facilitate discrimination between speaker voices. More details on this model can be found in \cite{diez2020optimizing}.

The configuration parameters of VB-HMM on x-vectors (VBx) were tuned on VoxConverse development set so that $F_A=0.3$, $F_B=16$ and $P_{loop}=0.9$. Table~\ref{tab:VB_performance} presents the comparison of results for AHC, AHC with undercluster threshold, and VBx initialized with the latter AHC labels (row 6). Note the more than 63\% relative improvement of VBx over the AHC result in terms of speaker error (miss and FA are defined by the VAD) and more than 8\% relative improvement in terms of JER. Table~\ref{tab:qty_speakers} presents the amount of recordings where the estimated number of speakers was less, equal or greater than the correct number. We see that VBx finds the correct number of speakers in more than two thirds of the files while AHC only does so in little more than one half.

\begin{table}[!tb]
\centering
\caption{Diarization performance on the development set.}
\label{tab:VB_performance}
\vspace{2mm}
\begin{tabular}{@{}
                  l
                  S[table-format=2.2]  
                  S[table-format=2.2]  
                  S[table-format=1.2] 
                  S[table-format=2.2]  
                  S[table-format=2.2] 
                  @{}}
  \toprule
        System       & 
        \multicolumn{1}{c}{DER} & 
        \multicolumn{1}{c}{Miss} & 
        \multicolumn{1}{c}{FA} &
        \multicolumn{1}{c}{Spk.} & 
        \multicolumn{1}{c}{JER} \\
  \midrule
Baseline (AHC)\footnotemark & 24.57 & 11.21 & 2.26 & 11.10 & 51.71 \\ 
\midrule
AHC & 5.73 & 3.10 & 0.47 & 2.16 & 21.56 \\ 
\hspace{0.3cm}+ reclustering & 5.47 & 3.10 & 0.47 & 1.90 & 21.40 \\ 
\midrule
AHC (u-cluster) & 7.23 & 3.10 & 0.47 & 3.66 & 20.20 \\  
\hspace{0.3cm}+ reclustering & 6.19 & 3.10 & 0.47 & 2.62 & 19.53 \\ 
\midrule
VBx & 4.36 & 3.10 & 0.47 & 0.79 & 19.78 \\ 
\hspace{0.3cm}+ reclustering & 4.30 & 3.10 & 0.47 & 0.73 & 19.81 \\ 
\bottomrule
\end{tabular}
\end{table}
\footnotetext{Note that these numbers differ from~\cite{chung2020spot}, as the authors scored with a different tool. All the numbers in this paper were produced with \url{https://github.com/nryant/dscore}, which was also the official scoring tool for the challenge.}

\vspace{-2mm}
\subsection{Reclustering}
\vspace{-1mm}
One of the main disadvantages of the proposed approach for diarization is that the embeddings are computed over short segments of speech. This is in part necessary due to the dynamic nature of conversational speech. However, if embeddings were extracted on longer segments where we have the belief that the whole segment belongs to the same speaker, then the embeddings might be more robust, leading to better clustering.

In this direction is that we propose a ``reclustering'' step where all segments of a speaker in the recording, given a previous diarization run, are concatenated to extract a new embedding. Then, the per speaker global x-vectors are clustered with AHC to join speakers if necessary. 

The application of the reclustering on different diarization outputs (obtained with AHC, AHC with the underclustering threshold and with VBx) are presented in Table~\ref{tab:VB_performance}. In all cases we applied PCA to rotate the space but keeping all dimensions. 
We see that for AHC, reclustering improves the speaker error by 12\% relative while for VBx it is around 7.5\% relative. It should be noted that the speaker error for VBx is already quite low before reclustering, possibly leaving less room for improvement in comparison with AHC.

\begin{table}[!tb]
\centering
\caption{Number of recordings where the amount of found speakers is greater, equal, and less than the correct amount (CA) and mean across recordings of the difference between the correct amount of speakers and the amount of found speakers.}
\label{tab:qty_speakers}
\vspace{2mm}
\setlength{\tabcolsep}{5pt} 
\begin{tabular}{@{}
                  l
                  S[table-format=3.0]  
                  S[table-format=3.0]  
                  S[table-format=2.0] 
                  S[table-format=1.2] 
                  @{}}
  \toprule
        System       & 
        \multicolumn{1}{c}{\#spk $>$ CA} & 
        \multicolumn{1}{c}{\#spk $=$ CA} & 
        \multicolumn{1}{c}{\#spk $<$ CA} &
        \multicolumn{1}{c}{mean} \\
  \midrule
Baseline (AHC) & 82 & 46 & 88 & 0.44 \\ 
\midrule
AHC & 67 & 110 & 39 & -0.50 \\ 
\hspace{0.3cm}+ reclustering & 67 & 110 & 39 & -0.42 \\ 
\midrule
AHC (u-cluster) & 143 & 55 & 18 & -3.09 \\ 
\hspace{0.3cm}+ reclustering & 140 & 56 & 20 & -2.63 \\ 
\midrule
VBx & 12 & 147 & 57 & 0.34 \\ 
\hspace{0.3cm}+ reclustering & 11 & 145 & 60 & 0.37 \\ 
\bottomrule
\end{tabular}
\end{table}

\vspace{-2mm}
\subsection{Overlapped speech handling}
\vspace{-1mm}
One common scenario in conversations with several participants is overlapped speech. In particular, in the VoxCeleb development set an average of 2.9\% of speech has two or more speakers speaking simultaneously~\cite{chung2020spot}. To illustrate the effect of overlapped speech on DER, if we had perfect VAD and correctly labeled all segments of speech with only one speaker (in segments with more than one we only mark one of the correct ones) we would still obtain 2.3\% missed speech. If we also labeled correctly the second speaker where at least two speakers speak simultaneously, we would obtain 0.08\% missed speech. Doing so for three speakers means already less than 0.01\% missed speech. In terms of JER, the results would be 5.38, 0.26 and 0.02 respectively.

Note that with a system that does not label segments for more than one speaker, even with perfect VAD and no speaker error, the DER can in the best case be 2.3\%. However, if a second speaker is correctly handled, that error can be decreased substantially. The advantage of modelling three or more speakers is negligible on these data and for this reason we focus only on labeling up to two speakers.

Since the pipeline up to this point outputs only one speaker per frame, a post-processing step is needed to add second labels. This requires doing OVD first, and then assigning a second speaker on the found segments. 

For overlap detection we used the model trained on AMI available in pyannote~\cite{Bredin2020} but we tuned the detection threshold on VoxConverse development in order to maximize the precision.
The performance on the development set is shown in Table~\ref{tab:ov_detection}. 

\begin{table}[!tb]
\centering
\caption{OVD performance on the development set in terms of frame accuracy, precision, recall, miss, and false alarm.}
\label{tab:ov_detection}
\vspace{2mm}
\begin{tabular}{@{}
                  l
                  S[table-format=2.2]  
                  S[table-format=1.3]  
                  S[table-format=1.3] 
                  S[table-format=1.2]  
                  S[table-format=1.2]  
                  S[table-format=1.2] 
                  @{}}
  \toprule
        OVD       & 
        \multicolumn{1}{c}{Acc.} & 
        \multicolumn{1}{c}{Prec.} & 
        \multicolumn{1}{c}{Reca.} &
        \multicolumn{1}{c}{Miss} & 
        \multicolumn{1}{c}{FA} & 
        \multicolumn{1}{c}{Error} \\
  \midrule
AMI-trained & 97.11 & 0.712 & 0.286 & 1.02 & 0.40 & 1.42 \\ 
\bottomrule
\end{tabular}
\end{table}

We evaluated two approaches for assigning a second speaker. An heuristic that considers the two closest speakers in time~\cite{otterson2007efficient} and, based on~\cite{Bullock2020}, an approach where the second most-likely speaker of the output of VB-HMM diarization is used to provide the second label, but applied using x-vectors as input frames instead of mel-frequency cepstral coefficients. Given the current pipeline, obtaining the second label is quite straightforward as we simply need to output the two most likely speakers for each frame.

Results comparing the OVD system and the oracle OVD are presented in Table~\ref{tab:ov_handling}, together with a comparison of the methods for selecting the second speaker: with VBx, with the heuristic or directly using the oracle labels.
While the second speaker from VBx provides a convenient and principled mechanism for selecting the second speaker in overlap segments, we see that the performance is similar or slightly worse than choosing the closest speaker in time. 
However, choosing the second speaker according to the oracle labels would still provide a notable gain as compared to any of these methods, showing that there is room for improvement in terms of overlap handling.

Nevertheless, oracle OVD proves to be significantly better than the OVD system. In this sense, there is far much more room for improvement on the detection of overlap rather than on the handling.
Still, it should be noted that overlapped speech labels on the development set, like in any other diarization dataset, are not perfect, as the labeling of these regions is a very challenging task even for human annotators, and the precision that can be achieved is hard to define.

\begin{table}[!tb]
\centering
\caption{Diarization performance on the development set for VBx + reclustering (without overlapped speech handling) and after doing OVD and handling. Performance is shown for the OVD system and the oracle OVD and for each case, the two handling approaches and the choice of oracle speaker are evaluated.}
\label{tab:ov_handling}
\vspace{2mm}
\begin{tabular}{@{}
                  l
                  S[table-format=1.2]  
                  S[table-format=1.2]  
                  S[table-format=1.2] 
                  S[table-format=1.2]  
                  S[table-format=2.2] 
                  @{}}
  \toprule
        System       & 
        \multicolumn{1}{c}{DER} & 
        \multicolumn{1}{c}{Miss} & 
        \multicolumn{1}{c}{FA} &
        \multicolumn{1}{c}{Spk.} & 
        \multicolumn{1}{c}{JER} \\
  \midrule
No handling & 4.30 & 3.10 & 0.47 & 0.73 & 19.81 \\ 
\midrule
System + VBx 2\textsuperscript{nd} & 4.02 & 2.36\footnotemark & 0.75 & 0.91 & 19.96 \\ 
System + heuristic & 4.00 & 2.34 & 0.75 & 0.91 & 19.80 \\ 
System + oracle & 3.56 & 2.36\footnotemark[6] & 0.47 & 0.73 & 18.78 \\ 
\midrule
Oracle + VBx 2\textsuperscript{nd} & 2.80 & 0.98\footnotemark[6] & 0.47 & 1.35 & 19.10 \\ 
Oracle + heuristic & 2.70 & 0.92 & 0.47 & 1.31 & 18.52 \\ 
Oracle + oracle & 2.20 & 1.00\footnotemark[6] & 0.47 & 0.73 & 16.17 \\ 
\bottomrule
\end{tabular}
\end{table}
\footnotetext{Note that if when reclustering the first and second most likely speakers from VBx or the first and second by oracle are merged, then there is no second speaker to select leading to an increase in miss.}

\vspace{-2mm}
\subsection{Ablation study with oracle labels}
\vspace{-1mm}
Given that the VAD and OVD play a big role in the final performance of our system, we decided to further study these two components.

Considering the VAD, we compare performance of our system using the full VAD as described in section \ref{sec:VAD}, the baseline VAD, the simple energy based VAD with short pauses removed and the oracle VAD. Results are presented in Table~\ref{tab:VAD_oracle}. Here we can directly see how the VAD error (see table \ref{tab:VAD_performance}) influences DER performance. The difference on VAD performance between the baseline and the simple energy VAD (6.71 error) translates in almost 12\% DER difference. The extra improvement using the full VAD model (2.36 error reduction) translates into only 1.39\% DER improvement. Finally the oracle VAD (virtually 2.98 better VAD total error) only brings 1.21\% DER improvement. It is worth noting that the boundaries of speech segments in the oracle labels are not perfect (for this reason 0.25\,s forgiveness collar is used in the challenge for computing DER) and the performance gains given by better VAD reduce in terms of DER as the VAD error approaches to zero.

Next, regarding the OVD we compare the OVD system and the oracle OVD and the speaker label assignment when using also oracle VAD. Results show how the OVD system provides only a small DER improvement while oracle OVD brings a large gain, around 50\% decrease in DER. The oracle assignment brings only an extra 15\% decrease in error showing again that the main room for improvement is in the overlap detection.

\begin{table}[!tb]
\centering
\caption{VAD error and diarization performance on the development set for different combinations of VAD (`O' refers to oracle) and OVD (`S' refers to the system and `O' refers to the oracle) with the heuristic. In all cases, VBx + reclustering system is used.}
\label{tab:VAD_oracle}
\vspace{2mm}
\setlength{\tabcolsep}{4.5pt} 
  \begin{tabular}{@{}
                  l
                  S[table-format=2.2]  
                  S[table-format=1.2] 
                  S[table-format=1.2]  
                  S[table-format=1.2] 
                  S[table-format=1.2]  
                  S[table-format=2.2] 
                  @{}}
  \toprule
        System       & 
        \multicolumn{1}{c}{\begin{tabular}{@{}c@{}}VAD \\ error\end{tabular}} &
        \multicolumn{1}{c}{DER} & 
        \multicolumn{1}{c}{Miss} & 
        \multicolumn{1}{c}{FA} &
        \multicolumn{1}{c}{Spk.} & 
        \multicolumn{1}{c}{JER} \\
  \midrule
  Full VAD & 2.98 & 4.30 & 3.10 & 0.47 & 0.73 & 19.81 \\ 
  \midrule
   Baseline VAD & 12.05 & 17.58 & 11.22 & 2.26 & 4.10 & 39.73 \\ 
   Energy VAD & 5.34 & 5.69 & 2.99 & 1.78 & 0.92 & 21.52 \\ 
   O VAD & 0 & 3.09 & 2.30 & 0 & 0.79 & 17.26 \\ 
   \midrule
   O VAD + S OVD h & 0 & 2.79 & 1.55 & 0.28 & 0.96 & 17.29 \\ 
   O VAD + O OVD h & 0 & 1.50 & 0.08 & 0 & 1.42 & 16.12 \\ 
   O VAD + O OVD O & 0 & 1.01 & 0.22 & 0 & 0.79 & 13.80 \\ 
   \bottomrule
  \end{tabular}
\end{table}

\vspace{-1mm}

\section{CONCLUSIONS}
\vspace{-2mm}
In this paper we have proposed a system for performing diarization on the new VoxConverse diarization corpus comprised by broadcast recordings. The pipeline consists of voice activity detection, embedding extraction, VBx using AHC as initialization, reclustering using per recording global embeddings and overlapped speech detection and handling. We have analyzed the effect of each step in the final performance and identified the aspects that need to be addressed in the future to improve the performance further.

Our best submission on the challenge, corresponding to DER 4\% and JER 19.80 on the development set, obtained on the evaluation set DER 8.12\% and JER 18.35 allowing us to obtain the second position in terms of DER and the first in terms of JER. Analyzing the reasons for such differences in results between development and evaluation remains for a post-evaluation stage once the evaluation labels are released. Together with this publication, the most relevant modules of our system are released to the public in a branch of~\cite{VBx}.

\bibliographystyle{IEEEbib}
\bibliography{refs}

\end{document}